\documentclass[12pt,useAMS,useAASTEX,usenatbib]{aastex}
\usepackage{emulateapj5}

\makeatletter

\newenvironment{apjemufigure}{%
\def\@captype{figure}%
\noindent\begin{minipage}{0.999\linewidth}\begin{center}}
{\end{center}\end{minipage}}
\makeatother

\def\etal{et al.}

\def\alm{a_{\ell m}}

\def\a{{\bf a}}

\newcommand{\nbi}{{Niels Bohr Institute, Blegdamsvej 17,
DK-2100 Copenhagen, Denmark}}

\newcommand{\ru}{{Rostov State University, Space Research Department,
Zorge,5, 344091, Russia}}
\newcommand{\asc}{{Astro Space Center of Lebedev Physical Institute,
Profsoyuznaya 84/32, Moscow, Russia}}

\slugcomment{Submitted to {\it The Astrophysical Journal Lett.}}

\begin{document}

\title{"4n multipole periodicity" of the Galaxy image in the WMAP data.}
 
\author{
Pavel D. Naselsky\altaffilmark{1,2},
Igor D. Novikov\altaffilmark{1,3}
}

\altaffiltext{1}{\nbi}
\altaffiltext{2}{\ru}
\altaffiltext{3}{\asc}

\email{naselsky@nbi.dk }


\begin{abstract}
We present a specific periodicity of the Galaxy images in the multipole domain which can be used
for separation of 
the cosmic microwave background (CMB) signal  
from the Galaxy and foreground  component. 
 This method takes into consideration all the 
 coefficients of the expansion of the signal from the sky $\Delta T(\Theta, \phi)$ into
  spherical harmonics and dealing with combination of multipoles $a_{l,m}$ and $a_{l+\Delta,m}$
 for each multipole mode ($\ell , -\ell \le m \le \ell$) from the
whole sky without galactic cut, masks or any dissection of the whole sky
into disjoint regions. For the polar coordinate system we use particular values for $\Delta=4n$, $n=1,2,...$
which remove all  bright point-like sources localized in the Galactic plane and strong diffused component
down to the CMB level. To
illustrate that significant correlations of the point like sources and foregrounds $a_{l,m},a_{l+\Delta,m}$ coefficients
we apply this method to the WMAP Q,V and W bands in order to remove these marks of Galaxy from the maps.
We believe that our method would be useful for  separation of the CMB signal from different
kind of "noises" in the maps. 
\end{abstract}

\section{Introduction}
   Separation of the CMB signal and foregrounds (especially, Galactic foregrounds) is the major problem for all the CMB experiments, 
including ongoing the WMAP and future the PLANCK missions. The basics approach to solve this problem is related
with implementation of different masks and disjoint regions of the whole sky maps (see,\cite{wmap},\cite{wmapmap}
\cite{toh},\cite{eriksen} ). However, the question is how the Galactic region of the map can affected
any estimators of the primordial CMB signal, particularly, the power spectrum $C(l)$ and phases of the multipole
representation of the CMB,
 if we  try to process the data sets without any masks? This question closely related to the question,
 what statistical properties of the Galactic foreground could be? 
 
 In this letter we would like to present some of the results of investigation of the Galactic area of the WMAP
 sky using the  $K-W$ bands signals. Particularly, we will show that in the polar coordinate system the
 brightest area of the map related with the Galaxy can be removed down to the CMB level by simplest 
 linear combination of the spherical harmonics coefficients $a_{l,m}$ 
 \begin{equation}
 d^{\Delta}_{l,m}=a_{l,m}-\frac{|a_{l,m}|}{|a_{l+\Delta,m}|}a_{l+\Delta,m}
 \label{eq1}
 \end {equation} 
where $\Delta $ is free parameter, which corresponds to transition from the $l$ multipoles to $l+\Delta$ multipoles
 for the same $|m|\le l$- modes.  
 As one can see from Eq. (\ref{eq1}) our idea is to introduce new characteristics of the $\Delta T(\theta, \phi)$
 field on the sphere  
  \begin{equation} 
  D^{\Delta}( \theta, \phi)=\sum_{l=2}^{l_{max}}\sum_{m=-l}^l d^{\Delta}_{l,m}Y_{l,m}( \theta, \phi)
\label{eq2}
 \end {equation}
(here $Y_{l,m}( \theta, \phi)$ are spherical harmonics) which can significantly decrease the Galactic plane signal in the K-W maps. 
In Section 1 we  show, that
this new estimator $d^{\Delta}_{l,m}$ directly related with phases and power spectrum of the whole sky signal and can be used
for separation of the CMB signal from foregrounds. Section 2 is devoted to practical implementation of the
method to the WMAP data sets. In this section we will show that for $d^{\Delta}_{l,m}$ with $\Delta=4n$,$n=1,2..$
the $D^{\Delta}( \theta, \phi)$ map derived from K-W maps  do not have strong signal from the Galaxy.
In Section 3 we discuss rotational invariance of $\Delta=4n$ correlations and show that they are specific
for particular orientation of the Galaxy in the reference system. In the conclusion we summarize all the results
and briefly discuss an application of the method for cleaning of the CMB maps from foregrounds.

\section{Properties of $d^{\Delta}_{l,m}$ estimator.}
To understand the properties of $d^{\Delta}_{l,m}$ estimator let us define the transformation of the
$\Delta T(\Theta, \phi)$ signal to spherical harmonic coefficients $a_{l,m}$ as 
\begin{equation}
\Delta T( \theta, \phi)=\sum_{l=2}^{l_{max}}\sum_{m=-l}^l |a_{l,m}|e^{i\Phi_{l,m}}Y_{l,m}( \theta, \phi)
\label{Esq3}
 \end {equation} 
where $|a_{l,m}|$ and $ \Phi_{l,m}$ are the magnitudes and phases of  $l,m$ harmonics correspondingly.
In terms of these variables our estimator $d^{\Delta}_{l,m}$ now has a form
\begin{equation}
d^{\Delta}_{l,m}=|a_{l,m}|e^{i\Phi_{l,m}}\left(1-\exp[i(\Phi_{l,m}-\Phi_{l+\Delta,m}]\right)
\label{eq4}
\end {equation} 
One can see, that $d^{\Delta}_{l,m}$ coefficients depend on the phase difference $\Phi_{l,m}-\Phi_{l+\Delta,m}$,
taking for $l, l+\Delta$ multipoles. If the phase difference is small enough $\Phi_{l,m}-\Phi_{l+\Delta,m}\ll \pi/2$, 
then 
\begin{equation}
d^{\Delta}_{l,m}\simeq|a_{l,m}|e^{i(\Phi_{l,m}+\frac{\pi}{2})}\left(\Phi_{l,m}-\Phi_{l+\Delta,m}\right)
\label{eq5}
\end {equation}  
 So, for Gaussian CMB signal the phase difference is not especially small because of non- correlation of the
 phases, while for non-Gaussian signal these correlation should be significant (see, for example, Chiang et al, 2003, Naselsky et al, 2004).
 According to Eq. (\ref{eq5}) 
 their contribution to $d^{\Delta}_{l,m}$ became to be vanish, if $\Phi_{l,m}$ is almost the same as $\Phi_{l+\Delta,m}$.
  
  To illustrate that tendency, let us describe the simplest model (see Naselsky et al, 2004), when a bright point sourse is placed in the
  Galactic plane area, having coordinates $\theta_j=\pi/2, \phi_j$ and amplitude $A_j$.
The phases of the $\alm^j$ harmonics for that signal are (Naselsky et al, 2004)
\begin{equation}
\tan\Psi_{l,m}=-\frac{\sin(m\phi_j)}{\cos(m\phi_j)}
\label{ps2}
\end {equation}
and they do not depend on $l$. So, for single point source $\Phi_{l,m}=\Phi_{l+\Delta,m}$ and that source
do not contribute to the $D^{\Delta}( \theta, \phi)$ map for all $\Delta=1,2,...$ in Eq(\ref{eq1}).
For combination of the sources $\Delta T(\theta,\phi)=\sum_j\Delta T_j(\theta,\phi)$ the properties of the phases
are more complex that for the single one (see Naselsky et al, 2004), but the tendency of "selfcleaning" of
the $D^{\Delta}( \theta, \phi)$ map should be preserved, at list for some values of $\Delta$-parameter.
In the next section we will show that for $\Delta=4n$, where $n=1,2..$ this method provide the best result for
the Galactic plane area, including not the point sources only, but diffused foreground as well. 

 \section{"4n" periodicity of the WMAP bands.}
 
 To investigate the properties of $d^{\Delta}_{l,m}$ estimator we use the HEALPix package (Gorski,Hivon \&
 Wandelt, 1999) to decompose each of the WMAP signal maps at K-W bands for the spherical harmonics coefficients
 $a^{K-W}_{l,m}$ ($N_{side}=512$). Then we
 perform transformation of the $a^{K-W}_{l,m}$ coefficients to $d^{\Delta}_{l,m}$ for each band and obtain the following
 maps, shown in Fig.\ref{fig1}. We reproduce the same analysis using the GLESP package (Doroshkevich et all, 2003)
 in order to show that corresponding properties of the maps do not depend on the sky pixelization .

To remove  contribution of the CMB signal to the $ D^{\Delta}( \theta, \phi)$ map, in Fig.\ref{fig2}
we plot $ D^{\Delta=4}( \theta, \phi)$ for difference between V and W maps, which containt the foregrounds and
instrumental noise only. 

In Fig.\ref{fig3} we show the $ D^{\Delta}( \theta, \phi)$ maps for W band and different $l_{max}$, but for the same
value $\Delta=4$.
 Note that for $l_{max}> 150$ the properties of $ D^{\Delta=4}( \theta, \phi)$ map depend on the instrumental noise,
  while for $l_{max}= 50$ all the $ D^{\Delta=4}( \theta, \phi)$ maps have quite regular structure, without
  Galactic belt. Important remark is that if $\Delta\neq 4n$ then $d^{\Delta}_{l,m}$ estimator does not remove the strong
  signal in the Galactic plane.

 \section{Rotational invariance of the $ D^{\Delta=4}( \theta, \phi)$ map .}
The $d^{\Delta=4}_{l,m}$   estimators of Galactic signal 
 are dealing with phases of $\a_{l,m}$ coefficients and consequently
depend on the reference system of coordinates . Obviously, for different system of coordinates these
$\a_{l,m}$ coefficients and corresponding phases are different for the same values  $l,m$.
 Practically speaking, are  $d^{\Delta=4}_{l,m}$ estimators
rotationally invariant and if not, how significantly that non-invariance can transforms any
conclusions about contribution of the Galactic signal? To answered to these questions we need to know
how to transform a given set of $a_{l,m}$ taking for given reference coordinate system to new coefficients
$b_{l,m}$ which corresponds to new coordinate system rotated by Euler angles $\alpha,\beta,
\gamma$. Follow to general method \cite{v} and taking 
into account properties
of spherical harmonics (see Coles ,2005, Naselsky et al, 2004) we get

\begin{equation} 
 b_{l,m}=\sum_{m^{'}} D^l_{m,m^{'}}(\alpha,\beta,\gamma)a_{l,m^{'}}
\label{rot1}
\end{equation}
where $D^l_{m,m^{'}}(\alpha,\beta,\gamma)$ is a spherical harmonic decomposition 
of the Wigner function $\textbf{D}(\alpha,\beta,\gamma)$. The coefficients 
$D^l_{m,m^{'}}(\alpha,\beta,\gamma)$ should preserve the modules $\sum_m|b_{l,m}|^2=
\sum_m|a_{l,m}|^2 $ under transformation. 
Without lost of generality let assumes that for initial reference system of coordinates 
$a_{l,m}=exp(i\Phi_{l,m})$ where $\Phi_{l,m}$ are the phases for given $l$ and $|m|\le l$.
To obtain the trigonometric moments in the reference system after rotation we define  a matrix of correlations
  \begin{apjemufigure}
\hbox{\hspace*{-0.5cm}
\centerline{\includegraphics[width=0.45\linewidth]{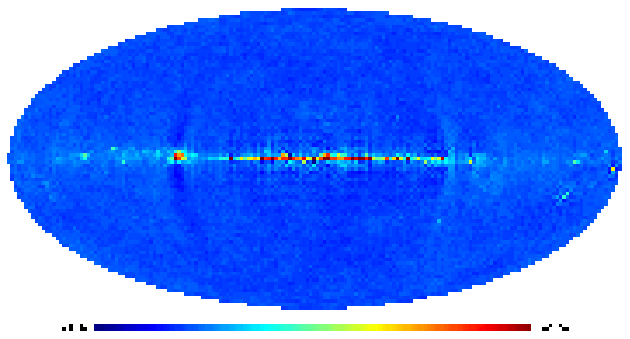}}}
\hbox{\hspace*{-0.5cm}
\centerline{\includegraphics[width=0.45\linewidth]{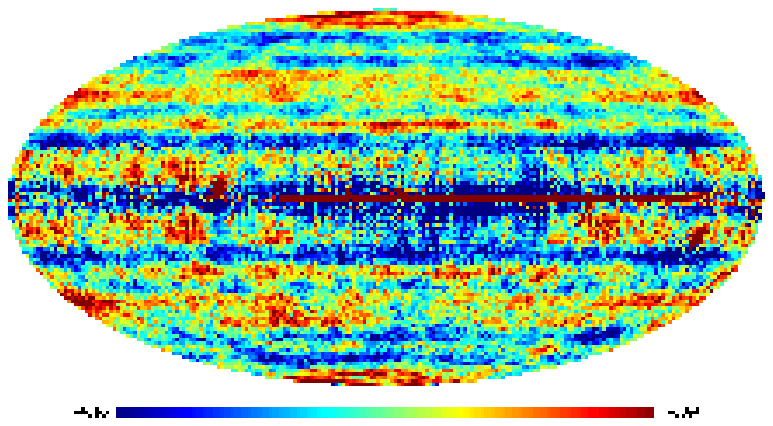}}}
\hbox{\hspace*{-0.5cm}
\centerline{\includegraphics[width=0.45\linewidth]{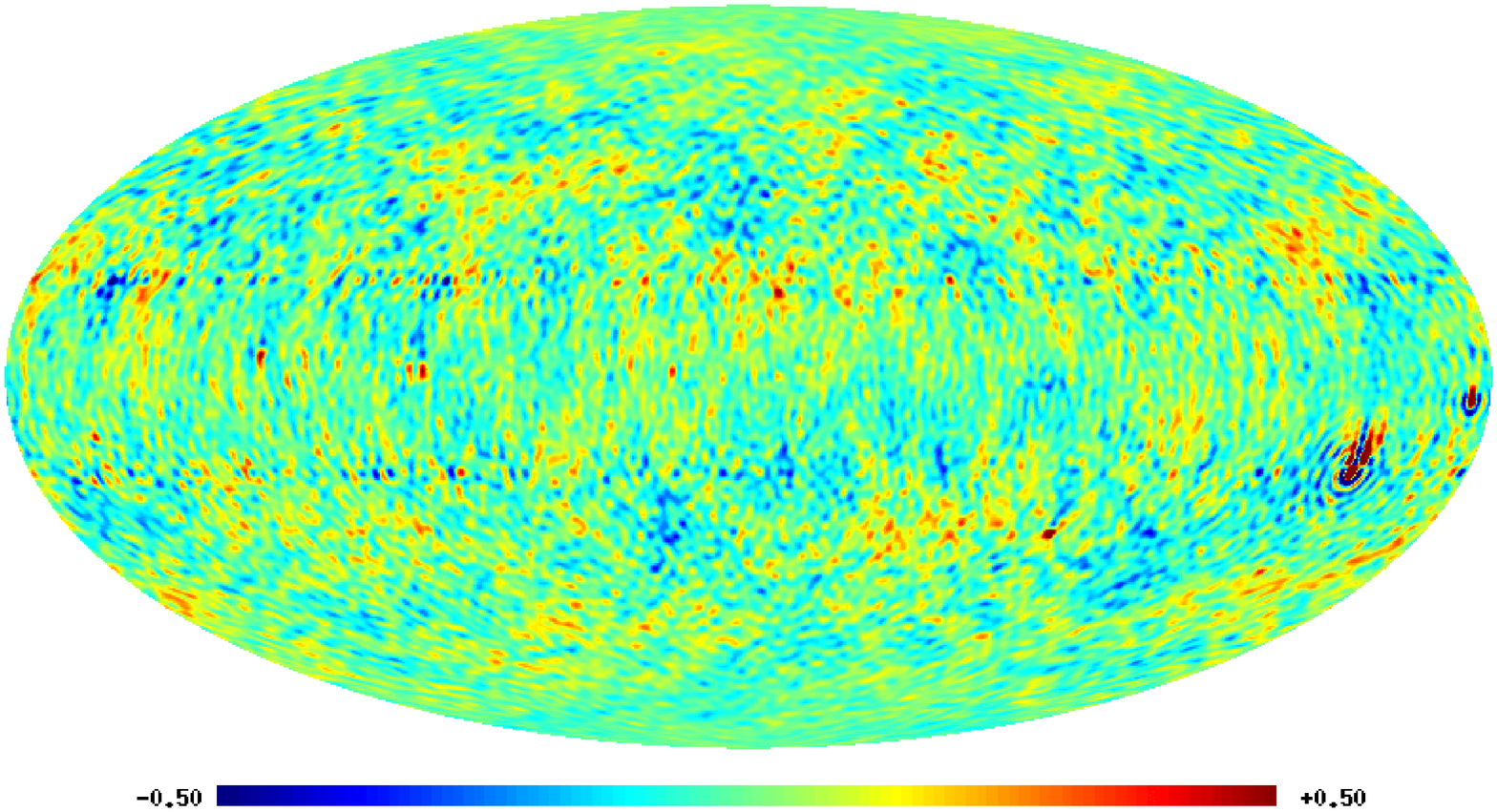}}}
\hbox{\hspace*{-0.5cm}
\centerline{\includegraphics[width=0.45\linewidth]{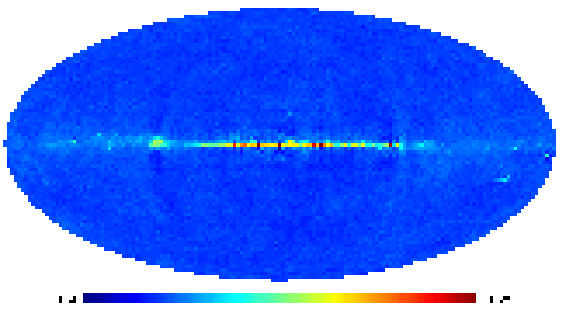}}}
\hbox{\hspace*{-0.5cm}
\centerline{\includegraphics[width=0.45\linewidth]{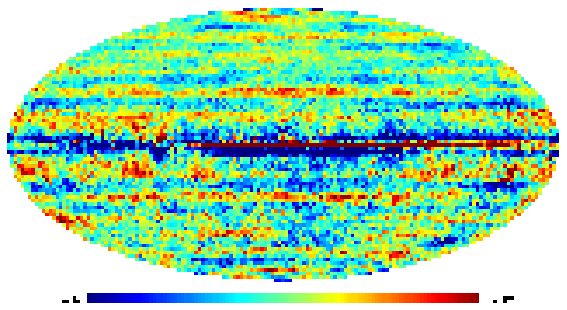}}}
\hbox{\hspace*{-0.5cm}
\centerline{\includegraphics[width=0.45\linewidth]{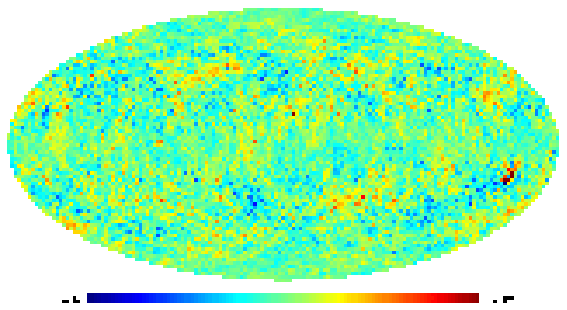}}}
\caption{From the top to the bottom : the $ D^{\Delta=4}( \theta, \phi)$ map for the  V band for $\Delta=0$ , $\Delta=1$ and $\Delta=4$.
Next tree maps are the same , but for W band. All the maps are taking for $l\le 250$. Here and bellow
the amplitudes of the signal are in micro Kelvin. For all $ D^{\Delta=4}( \theta, \phi)$ maps the
range of temperature scale is $-0.5,0.5 mK$.
 }  
\label{fig1} 
\end{apjemufigure}
 
\begin{eqnarray}
G_{l,l^{'},m,m^{'}}= b_{l,m}b^{*}_{l^{'},m^{'}}=\sum_{q,p} 
D^l_{m,q}(\alpha,\beta,\gamma)\nonumber\\
D^{*l^{'}}_{m^{'},p}(\alpha,\beta,\gamma)a_{l,q}a^{*}_{l^{'},p}
\end{eqnarray}
where $l^{'}=l+\Delta$, $m^{'}=m$ and $b_{l,m}=|b_{l,m}|exp(i\phi_{l,m})$, $ \phi_{l,m})$ is the phase of $l,m$
harmonics after rotation. 
 So, under rotation  the $ D^{\Delta=4}( \theta, \phi)$  map transforms by the following way:
\begin{eqnarray}
e^{i(\phi_{l,m}-\phi_{l^{'},m^{'}})}=\frac{e^{i\alpha(m-m^{'})}}{|b_{l,m}b^{*}_{l^{'},m^{'}}|}\nonumber\\
\sum_{p,q} D^l_{m,p}(\alpha,\beta,\gamma)
D^{*l^{'}}_{m^{'},q}(\alpha,\beta,\gamma)e^{i(\Phi_{l,p}-\Phi_{l^{'},q})}\nonumber\\
\label{rot5}
\end{eqnarray}

Taking into account that (see \cite{v})
\begin{equation}
D^l_{m,m^{'}}=\exp(-im\alpha)d^l_{m,m^{'}}(\beta)\exp(-im^{'}\gamma);\nonumber\\
\label{rot6}
\end{equation}
where
\begin{eqnarray}
d^l_{m,m^{'}}(\beta)=\left[(l+m^{'})!(l-m^{'})!(l+m)!(l-m)!\right]^{\frac{1}{2}}\times \nonumber\\
\sum_k^{min(l-m,l-m^{'})} (-1)^{l^{'}-m^{'}-k}\frac{(l-m-k)!k!}{(l-m^{'}-k)!(k+m+m^{'})!}\nonumber\\
\times(\cos\beta)^{2k+m+m^{'}}(\sin\beta)^{2l-2k-m-m^{'}}\nonumber\\
\label{rot7}
\end{eqnarray}
we obtain the following formula :

\begin{eqnarray}
e^{i(\phi_{l,m}-\phi_{l^{'},m^{'}})}=\frac{e^{i\alpha(m-m^{'})}}{|b_{l,m}b^{*}_{l^{'},m^{'}}|}\nonumber\\
 \sum_{p,q}W^{l,l^{'}}_{m,m^{'},p,q}(\beta)
e^{i\left(\Phi_{l,p}-\Phi_{l^{'},q}+\gamma(p-q)\right)}
\label{rot8}
\end{eqnarray}
where
$$
W^{l,l^{'}}_{m,m^{'},p,q}(\beta)= d^l_{m,p}(\beta)d^{l^{'}}_{m^{'},q}(\beta)
$$

The Eq(\ref{rot8}) show that after rotation the properties of the phase correlation became to be more complicated
than in the reference system. To estimate their analytically, we would like to point out that real and 
imagenary parts of the exponent in Eq. (\ref{rot8}) now containts not phase difference $\Phi_{l,p}-\Phi_{l^{'},q}$ only, but
additional terms $\gamma(p-q)$ and $\alpha(m-m^{'})$. However, taking into account that in the reference system
the major correlations corresponds to $q=p$ and $l^{'}=l+4n$, we can draw our attention on these modes only.
 As one can see from Eq. (\ref{rot8}), for these modes dependency of exponent in the left hand side on
 $\alpha$ and  $\gamma$ parameters is vanished   , while dependency
 on $\beta$ looks like a linear combination of exponents $\Phi_{l,p}-\Phi_{l+4n,p}$ convolved by the
 coefficients $W^{l,l+4n}_{m,m,p,p}(\beta)$. Obviously , that part of exponent is not rotationally invariant, 
 while rotation of the map with $\beta=0$ preserved rotationally invariance of the phase difference.
 To illustrate that effect in Fig.\ref{wrot} we plot the $D^{\Delta=4}( \theta, \phi)$ map taking from the
  W band, and rotated by the angles $\theta=77^o, \phi=38^o$ (which corresponds to $\beta \neq 0$) relatively to the system of standard polar coordinates.
  As one can see from this fig. the Galactic belt is still the major source of the high amplitude signal. 
  Thus, the effect of $4n$-periodicity is rotationally invariant for particular choosing of the Euler angles
  $\alpha\neq 0,\gamma\neq 0,\beta=0$.
 \begin{apjemufigure} 
\hbox{\hspace*{-0.5cm}
\centerline{\includegraphics[width=0.77\linewidth]{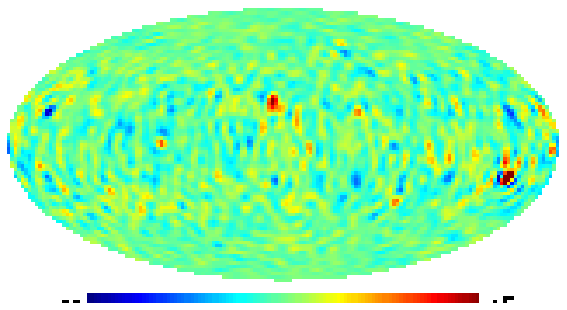}}}
\hbox{\hspace*{-0.5cm}
\centerline{\includegraphics[width=0.7\linewidth]{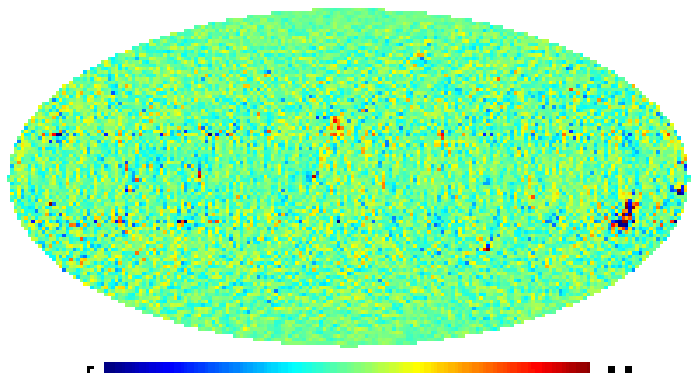}}}
\caption{ The maps $ D^{\Delta=4}( \theta, \phi)$  for the  difference  V -W bands 
for $\Delta=4$ and $l_{max}=50$(top). Bottom - the same map, but for $l_{max}=150$ 
}  
\label{fig2} 
\end{apjemufigure} 

\begin{apjemufigure}
\hbox{\hspace*{-0.5cm}

\centerline{\includegraphics[width=0.77\linewidth]{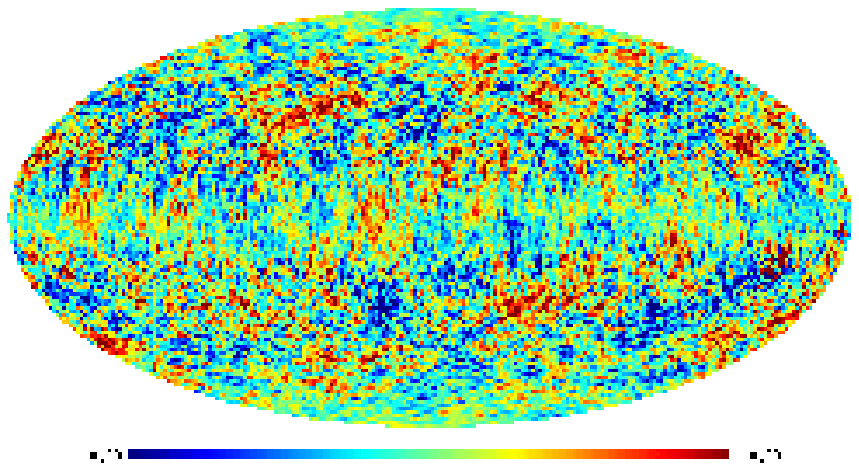}}}
\hbox{\hspace*{-0.5cm}
\centerline{\includegraphics[width=0.77\linewidth]{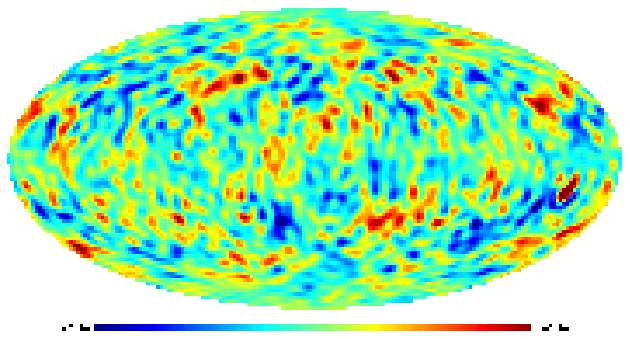}}}
\caption{From the top to the bottom : the $ D^{\Delta=4}( \theta, \phi)$ maps for W  band  for  
 $l_{max}=150$ and $l_{max}=50$.
 }  
\label{fig3} 
\end{apjemufigure}

 \begin{apjemufigure}
\hbox{\hspace*{0.01cm}
\centerline{\includegraphics[width=0.77\linewidth]{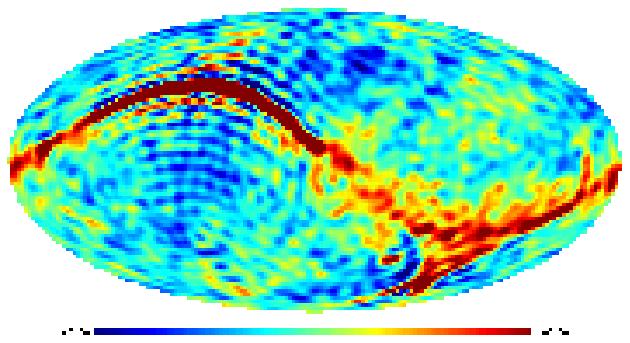}}}
\caption{The $D^{\Delta=4}( \theta, \phi)$ map  taking after rotation of the W band map by the angles 
$\theta=77^o, \phi=38^o$.
.}
\label{wrot} 
\end{apjemufigure}   
 At the end of this section we would like to point out that for given statistical properties of the
 primordial CMB signal the phases of  $d^{\Delta=4}_{l,m}$ estimator and power spectrum closely related to
 the phases $\xi_{l,m}$ and power spectrum of the signal $C(l)$, obtained using $\alm$ coefficients.
  For primordial Gaussian signal these
 relationships are extremely simple. The phases of $d^{\Delta=4}_{l,m}$ estimator should be non-correlated and
 uniformly distributed within interval $[0,2\pi]$, while the power spectrum 
 $D^{\Delta=4}(l)=1/(2l+1)\sum_m |d^{\Delta=4}_{l,m}|^2$ related to 
 standard one $C(l)=1/(2l+1)\sum_m |c_{l,m}|^2$ through equation

\begin{equation}
D^{\Delta=4}(l)=2C(l)-\frac{2}{2l+1}\sum_{m=-l}^{m=l}|c_{l,m}|^2\cos(\xi_{l+\Delta,m}-\xi_{l,m})
\label{pow} 
\end{equation}
 
From Eq. (\ref{pow}) one can see that $D^{\Delta=4}(l)\simeq 2C(l)$ with error about "cosmic variance" error. 
($\sim \frac{1}{\sqrt{l+\frac{1}{2}}}$).
 For non-Gaussian primordial CMB signal with correlated phases
 $\xi_{l+\Delta,m},\xi_{l,m}$ the power of $d^{\Delta=4}_{l,m}$ signal has a tendency to drop down to the level
 $D^{\Delta=4}(l)\rightarrow 0$, if $\xi_{l+\Delta,m}\rightarrow \xi_{l,m}$. That tendency open a new possibility
 to test Gaussianity of the CMB signal trough power spectrum estimation, using $\alm$ and $d^{\Delta=4}_{l,m}$
 coefficients and corresponding power spectra $C(l)$ and $D^{\Delta=4}(l)$. For Gaussian signal we should
 obtain $D^{\Delta=4}(l)\simeq 2C(l)$, while for non-Gaussian signal it should be $D^{\Delta=4}(l)< 2C(l)$ 
 with "cosmic variance" error bars.
 
 \section{Conclusion.}
 In this paper the new estimator for the CMB maps has been presented. This estimator is based on the phase
 correlation between $l+4n,m$ and $l,m$ harmonics and it decrease significantly contribution of the Galactic
 foregrounds to the map. Decrovered $4n, n=1,2...$ periodicity of the Galactic foegrounds phases seems to be
 important for accurate solution of the problem of separation the primordial CMB signal from the foregrounds and
 instrumental noise including whole sky signals. This method 
 will be publish in a separate paper.

 \section{Acknowledgments}
We thank Lung-Yih Chiang for useful discussions and help. We
acknowledge  the use of the Legacy Archive for Microwave Background
Data Analysis (LAMBDA) . We also
acknowledge the use of H{\sc ealpix} package \cite{healpix} to
produce $\alm$ and GLESP package \cite{glesp}



\begin{thebibliography}{}
\expandafter\ifx\csname natexlab\endcsname\relax\def\natexlab#1{#1}\fi

\newcommand{\combib}[3]{\bibitem[{#1}({#2})]{#3}} 

%
%
\newcommand{\autetal}[2]{{#1,\ #2., et al.,}}
\newcommand{\aut}[2]{{#1,\ #2.,}}
\newcommand{\saut}[2]{{#1,\ #2.,}}
\newcommand{\laut}[2]{{#1,\ #2.,}}

%
%
\newcommand{\refs}[6]{#5, #2, #3, #4} 
\newcommand{\unrefs}[6]{#5, #2 #3 #4 (#6)}  


%
%

\newcommand{\book}[6]{#5, #1, #2} 
%

\newcommand{\proceeding}[6]{#5, in #3, #4, #2} 




\def\apj{ApJ}
\def\apjl{ApJL}
\def\mn{MNRAS}  
\def\nature{Nature} 
\def\aa{A\&A}   
\def\prl{Phys.\ Rev.\ Lett.}
\def\prd{Phys.\ Rev.\ D}
\def\pr{Phys.\ Rep.}


\def\cambridgepress{Cambridge University Press, Cambridge, UK} 
\def\princetonpress{Princeton University Press}
\def\worldpress{World Scientific, Singapore}
\def\oxfordpress{Oxford University Press}




\combib{Bennett~et al.}{2003a}{wmap}
\autetal{Bennett}{C. L}
\refs{}
{\apj}
{583}
{1}
{2003}


\combib{Bennett~et al.}{2003b}{wmapmap}
\autetal{Bennett}{C. L}
\refs{}
{\apjs}
{148}
{1}
{2003}
{astro-ph/020320}

\combib{Bennett~et al.}{2003c}{wmapfg}
\autetal{Bennett}{C. L}
\refs{}
{\apjs}
{148}
{97}
{2003}
{astro-ph/0203208}


\combib{Coles}{2005}{col5}
\aut{Coles}{P}
{astro-ph/0502088}








\combib{Chiang, Naselsky, Verkhodanov \& Way}{2003}{tacng}
\aut{Chiang}{L.-Y} \aut{Naselsky}{P. D}
\aut{Verkhodanov}{O. V} \laut{Way}{M. J}
\refs{Non-Gaussianity of the derived maps from the first-year WMAP data}
{\apjl}
{590}
{65}
{2003}
{astro-ph/0303643}







\combib{Doroshkevich~\etal}{2003}{glesp}
\aut{Doroshkevich}{A. G} \aut{Naselsky}{P. D} \aut{Verkhodanov}{O. V}
\aut {Novikov}{D. I} \aut{Turchaninov}{V. I} \aut{Novikov}{I. D}
\laut {Christensen}{P. R}
\unrefs{}
{Int.J.of Mod.Phys}
{in press}
{}
{2005}
{astro-ph/0305537}




\combib{Eriksen~\etal}{2004}{eriksen}
\aut{Eriksen}{H. K} \aut{Hansen}{F. K} \aut{Banday}{A. J}
\aut{G\'orski}{K. M} \laut{Lilje}{P. B}
\refs{}
{\apj}
{612}
{633}
{2004}


\combib{G\'{o}rski, Hivon \& Wandelt}{1999}{healpix}
\aut{G\'{o}rski} {K. M} \aut{Hivon} {E} \laut{Wandelt} {B. D}
\refs{}
{Proceedings
of the MPA/ESO Cosmology Conference ``Evolution of Large-Scale
Structure'', eds. A. J. Banday, R. S. Sheth and L. Da Costa,
PrintPartners Ipskamp, NL}
{}
{}
{1999}
{}



\combib{Naselsky~\etal}{2004}{}
\aut{Naselsky}{P.}
\aut{Verkhodanov}{O}\aut{Olesen}{P} \laut{ Chiang}{ L.-Y.}
\refs{}
{\apj}
{615}
{45}
{2004}



\combib{Tegmark, de Oliveira-Costa \& Hamilton}{2003}{toh} 
\aut{Tegmark}{M}\aut{de Oliveira-Costa}{A} \laut{Hamilton}{A} 
\refs{}
{\prd}
{6813}
{3523}
{2003}

\combib{Varshalovich,Moskalev \& Khersonskii}{1988}{v}
\aut{Varshalovich}{D.A.} \aut{Moskalev}{A.V} \laut{ Khersonskii}{V.K} 
{1998}
{Quantum Theory of Angular Momentum. World Scientific, Singapure.}

\end{thebibliography}
\end{document}